\documentclass[11pt]{article}
\usepackage[margin=1.3in]{geometry}

\usepackage{times}
\usepackage{url}
\usepackage[small]{caption}
\usepackage{graphicx}
\usepackage{amsmath,amsfonts,mathtools}
\usepackage{booktabs}
\usepackage{algorithm}
\usepackage{algorithmic}
\usepackage{paralist}
\usepackage{color}
\usepackage{complexity}

\newtheorem{theorem}{Theorem}[section]
\newtheorem{example}{Example}[section]
\newtheorem{definition}{Definition}[section]

\newtheorem{corollary}{Corollary}[section]

\newcommand{\qed}{\unskip\hspace*{1em}\hspace{\fill}$\Box$}

\newcommand{\opt}{\mathsf{OPT}}
\newcommand{\alg}{\mathsf{ALG}}
\newcommand{\lb}{\mathsf{LB}}
\newcommand{\ub}{\mathsf{UB}}

\title{Defending with Shared Resources on a Network}

\author{Minming Li$^1$ \hspace{30pt} Long Tran-Thanh$^{2}$ \hspace{30pt} Xiaowei Wu$^{3}$\\
$^{1}$Department of Computer Science, City University of Hong Kong\\
\texttt{minming.li@cityu.edu.hk}\\
$^{2}$Department of Economics and Computer Science, University of Southampton\\
\texttt{l.tran-thanh@soton.ac.uk}\\
$^{3}$Faculty of Computer Science, University of Vienna, Austria\\
\texttt{xiaowei.wu@univie.ac.at}}

\begin{document}

\maketitle

\begin{abstract}
In this paper we consider a defending problem on a network.
In the model, the defender holds a total defending resource of $R$, which can be distributed to the nodes of the network. The defending resource allocated to a node can be shared by its neighbors. There is a weight associated with every edge that represents the efficiency defending resources are shared between neighboring nodes.
We consider the setting when each attack can affect not only the target node, but its neighbors as well. 
Assuming that nodes in the network have different treasures to defend and different defending requirements, the defender aims at allocating the defending resource to the nodes to minimize the loss due to attack.
We give polynomial time exact algorithms for two important special cases of the network defending problem.
For the case when an attack can only affect the target node, we present an LP-based exact algorithm.
For the case when defending resources cannot be shared, we present a max-flow-based exact algorithm.
We show that the general problem is \NP-hard, and we give a $2$-approximation algorithm based on LP-rounding.
Moreover, by giving a matching lower bound of $2$ on the integrality gap on the LP relaxation, we show that our rounding is tight. 
\end{abstract}

\section{Introduction}

In the recent years, security games have gained an increasing popularity within the artificial intelligence research community, and have been widely used in many areas of the field~\cite{sagt/LetchfordCM09,tambe2011security,aamas/YinT12}. 
Many of these games are played within a network structure (i.e., network security games), where  a defender protects a set of targets from an attacker by allocating defensive resources to nodes (or edges) of a network. Such problems include, but are not limited to, the following: designing network interdiction strategies for infectious disease control~\cite{assimakopoulos1987network}, cybersecurity mechanisms for defending computer networks~\cite{atal/SchlenkerTXFTTV18}, or police patrolling plans in urban security domains~\cite{ijcai/ZhangATWGJ17}.    
Existing network security models typically assume that: (i) one single security resource can be used to protect one single target only; (ii) the resource allocation happens in a binary manner, i.e., a target is either protected or not; and (iii) an attack on a single target does not have effect to other (possibly neighbouring) targets.  

However, in many real-world scenarios security resources often effectively protect multiple targets simultaneously. 
Furthermore, multiple resources can be allocated to the same target to strengthen the target's defence. 

\begin{example}
	Consider a police patrolling problem in which law enforcement forces are allocated to different districts of a city for patrolling. 
	The more resources are allocated to a certain district, the more successful crime prevention can be achieved. 
	In addition, in case of an urban crime event (e.g., bank robbery), patrolling forces from neighboring areas (i.e., nodes with edge connections to the target) can also provide help to the resources already at the target node.
\end{example}

On the other hand, successful attacks can also produce certain damages to neighboring targets.

\begin{example}
	Consider the scenario when a chemical terrorist attack in which a toxic chemical weapon (e.g., sarin gas, or nerve agents) is used in a certain area. As the wind can blow the gas away from the original target area, the weapon can also generate damage in surrounding areas (although this damage is typically weakened, compared to the damage the attack would achieve at the original target).
\end{example}

Put differently, in many real-world applications, the damage depends not only on how well the node under attack is defended, but also on the defence of its neighbors.
As these examples demonstrate, it is essential to take into account both the ability of sharing defending resources between nodes of the network, and the wide coverage of the attacks.
In this paper, we consider a general defending problem on a network where
\begin{itemize}
	\item the defending resource allocated to a node can be shared by its neighbors;
	\item the damage due to attack at a target node depends not only on the defending power of the target node, but also on that of its neighboring nodes.
\end{itemize}

\subsection{Related Work}
As mentioned above, most of the existing work in the security domain ignore resource sharing between nodes. A notable exception is the work of~\cite{aaai/GanAV15}, in which allocating a defending resource to a node can also protect the neighbors of that node. 
However, their model only looks at the binary version of resource allocation, where allocating multiple resources to the same node is not considered. Thus it can not be used to tackle our problem.
To address the multi-resource allocation problem,~\cite{ijcai/VuLS18} has looked at a Colonel Blotto formulation of the security game setting. However, their work does not exploit the underlying network structure. In addition, they do not consider the effect of the attacks to the surrounding nodes. 
It is worth noting that there is a line of security games research that look at attackers with multiple resources~\cite{ijcai/KorzhykCP11,ijcai/YinVAH16,sigmetrics/WangS17}. These models can be seen as somewhat relevant to our work as a single attack can have impact to more than one nodes. 
In addition,~\cite{games/ChanCO17} looked at the case that having an insufficiently protected node can affect the defence level of neighboring nodes, which is similar to our setting. However, these models do not take into account the defence side network-based resource sharing. 

There is a large body of literature that studies contagion in network security games. 
For example,~\cite{bachrach2013contagion,acemoglu2016network,lou2017multidefender} looked at stochastic contagion in network security problems. 
However, their model assume that the contagion is independently decided at each node, which is not the case in our setting.
In addition, \cite{aaai/TsaiNT12,alshamsi2018optimal} studied a shared resource model in which two players, namely the attacker and the defender, try to maximize their influence on a network.
Similary,~\cite{nguyen2009stochastic,vorobeychik2015securing} looked at generic target interdependency (i.e., an attack at one target might affect other targets as well) models.
However, these models do not discuss the defending thresholds or the loss due to attack, and thus are different from our model.

\subsection{Our Results}

Against this background, this paper addresses the network defending problem with shared resources in the following way: 
To capture the resource sharing ability, we allow a node $v$ to share its resource to its neighbor $u$ (weighted with a certain sharing coefficient $w_{uv}$).
In addition, we assign two defence level thresholds $\lb_u \leq \ub_u$ to each node $u$ to represent the spatial spread effect of an attack as follows: at each target node $u$,  we need at least $\lb_u$ resources to prevent any damages at the local level (i.e., on node $u$), and we need at least $\ub_u$ resources to stop the spread of the attack to neighboring nodes of $u$ (for more details see Section 2). 

Given this model, we first look at two special cases, namely: (i) when an attack cannot spread to the target node's neighbours; and (ii) when sharing defending resources is not feasible. 
The former can be captured by setting $\lb_u = \ub_u$ for each node $u$, and thus, we refer to it as the Single Threshold Model. The latter is referred to as the Isolated Model (as resource sharing is not allowed between neighbors)\footnote{The name ``isolated'' means that defending resources can not be shared. However, the damage due to attack still depends on the defending powers of the target node and its neighbors.}.

In particular, we prove the following theorems:

\begin{theorem}[Single Threshold Model]\label{th:discountless}
	The single threshold network defending problem can be solved in $O(n^\omega \log n)$ time, where $n$ is the number of nodes in the network and $\omega\approx 2.373$ is the matrix multiplication factor.
\end{theorem}

\begin{theorem}[Isolated Model]\label{th:isolated}
	The isolated network defending problem can be solved exactly in $O(mn\log n)$ time, where $n$ and $m$ are the number of nodes and edges in the network, respectively.
\end{theorem}

We also show that the general case of the problem is \NP-hard, and therefore, we propose a $2$-approximation algorithm. In particular, we prove the following theorems.

\begin{theorem}[Problem Hardness]\label{th:hardness}
	The network defending problem is \NP-hard.
\end{theorem}

\begin{theorem}[Approximation Algorithm]\label{th:approx}
	There exists a $2$-approximation algorithm for the network defending problem that runs in $O(mn\log n)$ time ($n$ is the number of nodes, and $m$ is the number of edges).
\end{theorem}

We remark that our algorithm approximates the problem in a resource augmentation manner.
That is, we show that by using a total defending resource $2R$, the object of our algorithm is at most that of any algorithm that uses defending resource $R$.
As we will show in Section~\ref{sec:hardness}, the problem with the objective of minimizing the damage does not admit any polynomial time approximation algorithm, unless \P=\NP.

\section{Model Description}

We model the network as an undirected connected graph $G(V,E)$, where each node $u\in V$ has a lower bound $\lb_u$ and an upper bound $\ub_u$, where $\lb_u\leq \ub_u$, that represent the defending requirement.
Besides, each node $u$ has a value $g_u$ and a discounted value $g'_u \leq g_u$ that represent the damage due to attack on node $u$.

\begin{definition}[Defending Resource and Defending Power]
	The defender has a total resource of $R$ that can be distributed to nodes in $V$, where $r_u$ is the \emph{defending resource}\footnote{In our model the resource can be allocated continuously.} allocated to node $u$, and $\sum_{u\in V}r_u = R$.
	
	The \emph{defending power} of node $u$ is given by
	\begin{equation*}
	\textstyle p_u := r_u+\sum_{v:(u,v)\in E} w_{uv}\cdot r_v,
	\end{equation*}
	where $w_{uv}$ is the \emph{weight} of edge $(u,v)$ that represents the efficiency defending resource is shared between $u$ and $v$.
\end{definition}

When the attacker attacks a node $u$:
\begin{enumerate}
	\item If the defending power $p_u \geq \ub_u$, the attacker gains $0$.
	\item If the defending power $p_u \in [\lb_u, \ub_u)$, the attacker gains $g'_u$ if $u$ has neighbor $v$ with defending power $p_v < \lb_v$; gains $0$ otherwise.
	\item If the defending power $p_u < \lb_u$, the attacker gains $g_u$.
\end{enumerate}

The intuition behind this formulation is that if there is sufficiently large defending power at target node $u$ (i.e., $p_u \geq \ub_u$), the attack can be quickly mitigated (e.g., the robber will be quickly caught, or the toxic gas can be completely neutralized).
Thus neither the target node nor its neighbors suffer from any damage.
On the other hand, if the attack cannot be quickly mitigated, but the defending power at target $u$ is sufficient to locally stop the attack (i.e., $\lb_u \leq p_u < \ub_u$), then the attack may spread to the neighbors of $u$ with a weakened power. In this case, the weakened attack achieves some success (i.e., $g'_u$  damage) if some neighbor of $u$ is not sufficiently protected (i.e., has inadequate defending power).
Finally, if the target node itself has insufficient protection (i.e., $p_u < \lb_u$, the attack achieves its maximal damage $g_u$.

The objective of the defender is to allocate the defending resource to the nodes to minimize the gain of the attacker (which can attack only one node).
We call $\{r_u\}_{u\in V}$ a \emph{defending strategy}.
For every defending strategy, there exists a node by attacking which the attacker gains the most.
We call the resulting gain of the attacker the \emph{defending result}.

We call the model \emph{single threshold} if $\lb_u = \ub_u$ for all nodes $u \in V$: in this model, when some node $u$ is attacked, the attacker gains either $g_u$ or $0$, e.g., the attack does not spread to any neighbor of $u$.
This model is inspired by many real-world scenarios, ranging from urban crime and conventional terrorist attacks, to various cybersecurity threats and wildlife reservation problems (i.e., green security games). The common in these scenarios is that a single attack does not have a spreading effect, and thus the damage does not depend on the defending power of neighboring nodes. 

We call the model \emph{isolated} if $w_{uv}=0$ for all edges $(u,v)\in E$.
The intuition behind this special case is that in some real-world applications, we cannot share resources between nodes. 
For example, consider a disease outbreak scenario where the success of defence depends on the size of vaccinated population at each region. While the disease itself can spread to the neighboring areas if the vaccinated population is not sufficiently high (i.e., $p_u < \ub_u$), the strength of defence of a particular area cannot be transferred to another region. 

We use $N(u):= \{v: (u,v)\in E\}$ to denote the set of neighbors for every node $u\in V$.
We use $n$ and $m$ to denote the number of nodes and edges in the graph $G$, respectively.
We use $\opt$ to denote the optimal defending result.

\section{Exact Algorithms for Special Cases}

Before turning to the general version of the problem, as a warm-up, we first consider the two special cases in this section.
We present polynomial time algorithms that solve the two cases exactly.

\subsection{Na\"{i}ve Attempt}

\begin{figure}[h]
	\centering
	\includegraphics*[width=0.45\textwidth]{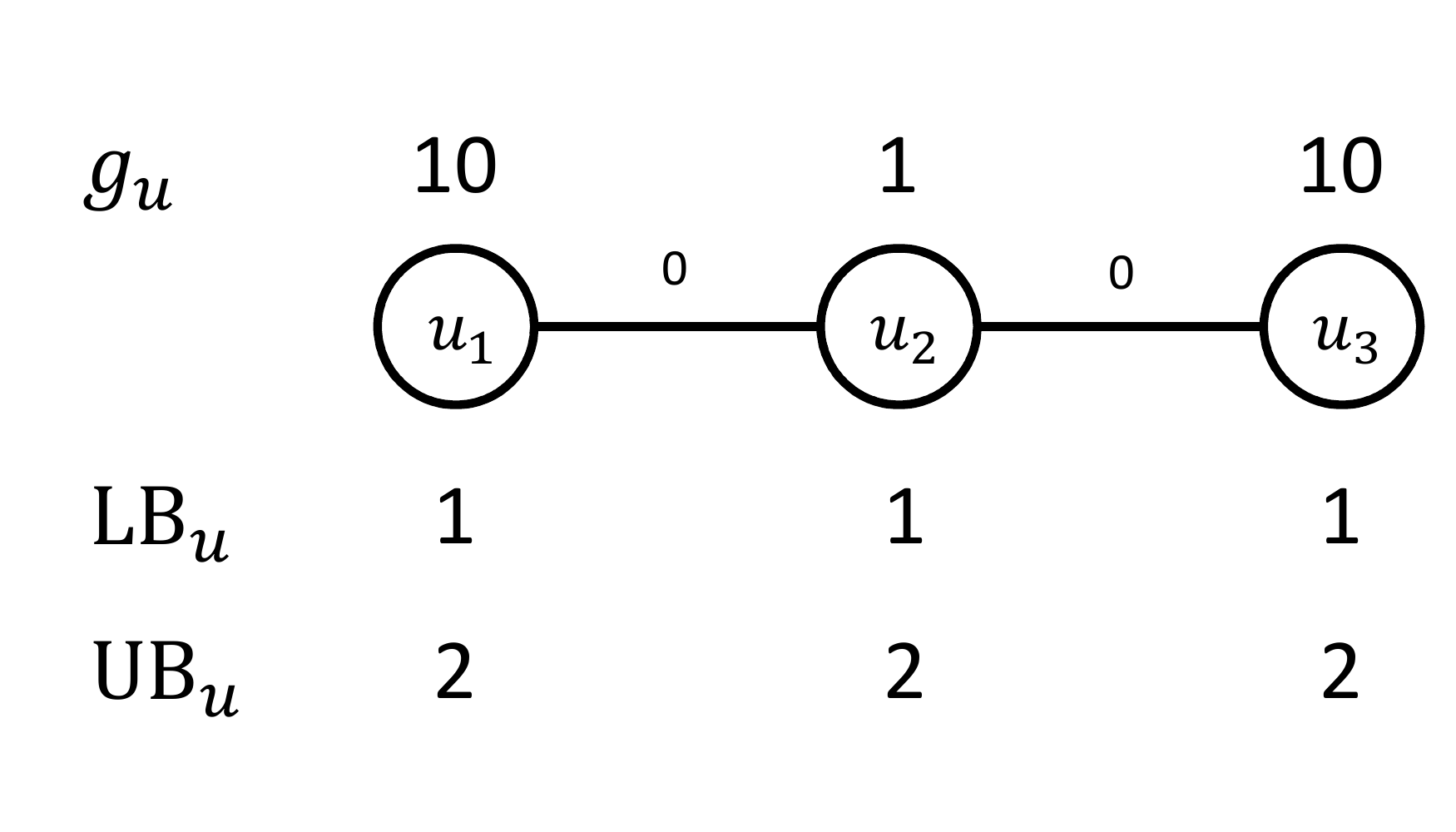}
	\caption{Hard instance for Greedy, in which $g'_u = g_u$, $R = 3$, and $w_{u_1 u_2} = w_{u_2 u_3} = 0$ (isolated model).
		It is easy to check that for this instance, $\opt = 0$ by allocating one unit of resource to each node, while $\alg = 10$.}
	\label{fig:greedy}
\end{figure}

Observe that for any defending strategy, there exists a node on which the attacker has maximum gain, which we refer to as the \emph{vulnerable} node.
Since the goal of the problem is to minimize the defending result, a natural algorithm would keep allocating defending resources to the vulnerable node until all resources are spent.
We refer to this algorithm as Greedy.
Unfortunately, as Figure~\ref{fig:greedy} shows, even in the isolated model, Greedy can perform arbitrarily bad.
Since node $u_1$ has the largest value, the Greedy algorithm will allocate at least 2 units of resource to $u_1$.
Consequently, both $u_2,u_3$ have defending power at most $1$, which results in $\alg = 10$, where $\alg$ is the defending result of the Greedy algorithm.

The intuition is, there are two solutions to protect the vulnerable node $u$ in the isolated model by allocating defending resources: either we guarantee that $p_u\geq \ub_u$, or all neighbors $v$ of $u$ has $p_v\geq \lb_v$.
However, solving the problem ``locally'' (as Greedy does) does not lead to a good defending strategy. 
This observation implies that the defending resources should be allocated in a ``global'' way that considers the effect on both the vulnerable node and its neighbors.

While this problem does not exist in the single threshold model, it is easy to show that the Greedy also performs arbitrarily bad in this model: consider the graph instance shown in Figure~\ref{fig:greedy}, where we change $w_{u_1 u_2} = w_{u_2 u_3} = 1$ and $\lb_u = \ub_u = 3$ for all $u\in \{u_1,u_2,u_3\}$.
Obviously we still have $\opt = 0$.
However, Greedy will allocate all resources to either $u_1$ or $u_3$, which results in $\alg = 10$.

The two hard instances for the Greedy algorithm imply that to solve the problem, we need to take into account the effect on neighbors of $u$, when allocating defending resource to a node $u$.
A key difference between our algorithms and the Greedy algorithm is that we set a defending result goal before allocating any resource, and try to allocate resource globally to achieve this goal.
In order to produce defending strategies in a global manner, we use the linear programming (LP) and maximum flow techniques.

\subsection{Single Threshold Model}\label{ssec:poly_lb=ub}

We first consider the single threshold model, i.e., $\lb_u = \ub_u$ for every node $u\in V$.
We show that combining the linear programming technique with a simple binary search, we can solve this problem exactly in polynomial time.

Since $\lb_u = \ub_u$ for all $u\in V$, the attack is either successful or unsuccessful immediately after attacking some node.
Therefore, while there are infinitely many defending strategies, the number of defending results is bounded by $n+1$.
Let $\mathcal{G} = \{ g_u: u\in V \}\cap \{ 0 \}$ be the defending result space, i.e., the possible defending results.

Suppose for every $\alpha\in \mathcal{G}$, we can decide in polynomial time whether the defending result $\alpha$ is \emph{achievable} (and output a solution if it is), then we can compute the optimal defending strategy in polynomial time. 
Here we call $\alpha\in\mathcal{G}$ achievable if and only if there exists a defending strategy (using $R$ defending resource) with defending result $\alpha$.
In the following, we show how to decide the achievability for every $\alpha\in\mathcal{G}$.
More importantly, we output a feasible defending strategy if $\alpha$ is achievable.

\begin{definition}[Vulnerable Nodes]
	Let
	\begin{equation*}
	A_\alpha := \{ u\in V:g_u > \alpha \}
	\end{equation*}
	be the nodes that need a defending power at least $\lb_u$, if the target defending result is $\alpha$.
\end{definition}

By definition, if any of $u\in A$ has defending power $p_u < \lb_u$, then the attacker gains $g_u > \alpha$ by attacking $u$, which violates our target defending result.
Hence the goal is to compute a defending strategy under which every node $u\in A$ has defending power at least $\lb_u$.

This is actually a simple task, as we can formulate the problem as a feasibility linear program, in which the defending resources allocated to the nodes are the variables.
\begin{align*}
\min. \qquad\qquad 0&\\
\text{s.t.}\qquad \textstyle \sum_{u\in V} r_u & = R,\\
\textstyle r_u + \sum_{v\in N(u)}w_{uv}r_v & \geq \lb_u, \quad \forall u\in A_\alpha \\
r_u & \geq 0, \qquad \forall u\in V.
\end{align*}

Observe that any feasible solution for the above LP gives a defending strategy with defending result at most $\alpha$.
On the other hand, if the LP is infeasible, then the target defending result $\alpha$ is not achievable.

By the state-of-the-art result by~\cite{stoc/CohenLS19}, we can solve the above LP in $O(n^\omega)$ time, where $\omega\approx 2.373$ is the matrix multiplication factor.
Thus by trivially enumerating all possible values of $\alpha\in \mathcal{G}$, we can solve the problem in $O(n^{\omega+1})$ time.
Indeed, observe that if $\alpha$ is achievable, then all values at least $\alpha$ are also achievable.
Thus by using a binary search on $\alpha\in\mathcal{G}$, we can output the feasible solution for the minimum achievable $\alpha$ as the optimal defending strategy in $O(n^\omega \log n)$ time, which completes the proof of Theorem~\ref{th:discountless}.

\subsection{Isolated Model}\label{ssec:poly_isolated}

Next we turn to the isolated model.
Recall that by definition we have $w_{uv}=0$ for all $(u,v)\in E$, but nodes can have arbitrary upper and lower bounds.
We show in this section that the isolated model can be solved exactly in polynomial time, using the maximum flow technique.
Similar to the previous analysis, observe that given $g_u$ and $g'_u$ for all $u\in V$ and $c$, we have $\opt \in \mathcal{G}:=  \{g_u:u\in V\}\cup \{g'_u:u\in V\} \cup \{0\}$. In other words, there are at most $2n+1$ different defending results.
Hence to solve the problem, we only need to identify the minimum value in $\mathcal{G}$ that is achievable using $R$ defending resource, and output a defending strategy achieving it.

Let $\alpha \in  \mathcal{G}$ be the aim of the defending result.

\paragraph{Algorithm for Testing the Achievability of $\alpha$.}
As before, we define $A_\alpha=\{ u\in V:g_u > \alpha \}$ to be the vulnerable nodes that need to receive at least $\lb_u$ of defending power.
Note that in the isolated model, we must allocate a defending resource at least $\lb_u$ to every $u\in A_\alpha$ to guarantee a defending result at most $\alpha$.
Let $B_\alpha\subseteq A_\alpha$ be $\{u\in V: g'_u> \alpha\}$, i.e., those $u\in A_\alpha$ who need to receive at least $\ub_u$ of defending power, or each of its neighbors $v\in N(u)$ has $p_v\geq \lb_v$.
We call $B_\alpha$ the set of \emph{crucial} nodes.
It remains to decide which crucial nodes $u$ should be assigned defending power $\ub_u$ (the remaining nodes will be covered by their neighbors).
For ease of notation we drop the subscript $\alpha$ on $A_\alpha$ and $B_\alpha$ in the following discussion.

Suppose $S\subseteq B$ are the nodes we decide to allocate a defending resource of $\ub_u$ to, i.e., we allocate an extra $\ub_u - \lb_u$ to every node $u\in S$.
To achieve the defending result, we need to guarantee that every neighbor $v\in N(u)$ of $u\in B \setminus S$ receives at least $\lb_v$.
Since all nodes $u\in A$ are already allocated a defending resource $\lb_u$, it suffices to allocate $\lb_v$ to every $v\in (V\setminus A)\cap  \bigcup_{u\in B\setminus S} N(u)$.

Let $N(B\setminus S) = \bigcup_{u\in B\setminus S} N(u)$ be the neighbors of nodes in $B\setminus S$.
The total resource required is given by
\begin{equation*}
\sum_{u\in A} \lb_u + \sum_{u\in S} (\ub_u-\lb_u) + \sum_{v\in (V\setminus A)\cap N(B\setminus S)} \lb_v.
\end{equation*}

In the following, we show that the problem of computing the set $S\subseteq B$ that minimizes the total defending resource can be solved by computing a maximum flow on a directed network with $O(|V|)$ nodes and $O(|E|)$ edges.
Note that the defending result $\alpha$ is achievable if and only if the minimum defending resource required is at most $R$.

\paragraph{Flow Network.}
Let the nodes of the directed flow network be $\{ u_\text{in},u_\text{out} : u\in B \}\cup \{ v_\text{in},v_\text{out} : v\in V\setminus A \}\cup \{s,t\}$, where $s$ is the source and $t$ is the sink.
The edges of the network are constructed as follows.
\begin{enumerate}
	\item 	For every $u\in B\cup (V\setminus A)$, let there be a directed edge from $u_\text{in}$ to $u_\text{out}$.
	For all $u\in B$, let the capacity of the edge be $\ub_u - \lb_u$.
	For all $v\in V\setminus A$, let the capacity of the edge be $\lb_v$.
	
	\item For every $u\in B$, let there be a directed edge from $s$ to $u_\text{in}$.
	For every $v\in V\setminus A$, let there be a directed edge from $v_\text{out}$ to $t$.
	For every $(u,v)\in E$ such that $u\in B$ and $v\in V\setminus A$, let there be a directed edge from $u_\text{out}$ to $v_\text{in}$.
	Let the capacity of these edges be infinity.
\end{enumerate}

\begin{figure}[h]
	\centering
	\includegraphics*[width=0.7\textwidth]{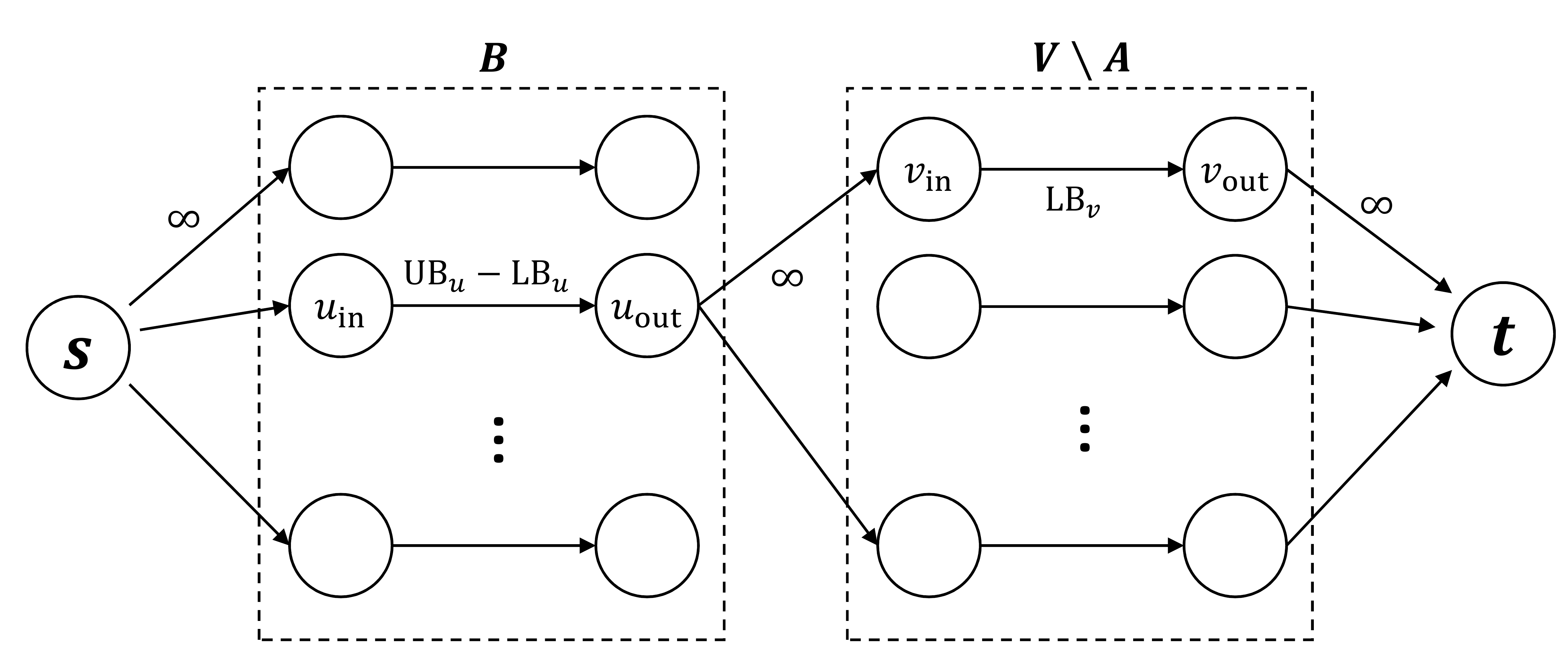}
	\caption{Illustrating figure of the flow network.}
\end{figure}

The flow network has $O(n)$ nodes and $O(n+m)$ edges.

Intuitively, we construct a directed network based on the bipartite graph between $B$ and $V\setminus A$.
Observe that in the directed flow network, to separate $s$ and $t$, either the edge $(u_\text{in},u_\text{out})$ is cut for $u\in B$, or $(v_\text{in},v_\text{out})$ is cut for every neighbor $v\in V\setminus A$ of $u$.
By setting the capacities as above, we guarantee that every cut separating $s$ and $t$ corresponds to a feasible defending strategy.

By the max-flow min-cut theorem, computing the maximum flow from $s$ to $t$ is equivalent to finding the minimum $s$-$t$ cut.
Consider the minimum $s$-$t$ cut that partitions the nodes into two sets $S$ and $T$, such that $s\in S$ and $t\in T$.
Let $\text{cut}(S,T)$ be the total capacity of edges between $S$ and $T$.

Since the edges from $s$ and the edges to $t$ have infinite capacity, we have $u_\text{in}\in S$ for all $u\in B$, and $v_\text{out}\in T$ for all $v \in V\setminus A$.
Observe that for every $u\in B$,
\begin{enumerate}
	\item if $u_\text{out} \notin S$, then the edge $(u_\text{in},u_\text{out})$ is cut by $(S,T)$;
	
	\item if $u_\text{out} \in S$, then for every $v\in N(u)\cap (V\setminus A)$, i.e., neighbor of $u$ that is not in $A$, we must have $v_\text{in} \in S$, as the capacity of the edge from $u_\text{out}$ to $v_\text{in}$ is infinity.
	Consequently, we know that edge $(v_\text{in},v_\text{out})$ is cut by $(S,T)$.
\end{enumerate}

Let $B_s := \{ u\in B: u_\text{out} \in S \}$ and $B_t = B\setminus B_s$.
Let $N(B_s) = \bigcup_{u\in B_s} N(u)$ be the neighbors of nodes in $B_s$.
Then we have
\begin{equation*}
\text{cut}(S,T) = \sum_{u\in B_t} (\ub_u - \lb_u) + \sum_{v\in N(B_s)\cap (V\setminus A)}\lb_v .
\end{equation*}

In other words, the total capacity of the cut is exactly the defending resource required to increase the defending power of every $u\in B_t$ from $\lb_u$ to $\ub_u$, and every neighbor $v$ of $u\in B_s$ that is not in $A$ from $0$ to $\lb_v$.
Hence the minimum $s$-$t$ cut corresponds to the optimal defending strategy with defending result $\alpha$.

\paragraph{Running Time.}
The maximum flow problem on a directed network with $n$ nodes and $m$ edges can be solved in $O(mn)$ time by~\cite{stoc/Orlin13}. Testing the achievability of every $\alpha\in \mathcal{G}$ (and outputting a solution, if any) can be done in $O(mn)$ time.
As before, by adopting a binary search on values of $\mathcal{G}$, we can identify the minimum achievable $\alpha$ in $O(mn\log n)$ time, which completes the proof of Theorem~\ref{th:isolated}.

\section{Hardness Results}\label{sec:hardness}

As we have shown in the previous section, for the single threshold model (i.e., every node $u$ has $\lb_u = \ub_u$) and the isolated model (i.e., $w_{uv}=0$ for all $(u,v)\in E$), the problem is polynomial time solvable.
A natural idea is to combine the two techniques we have used to solve the special cases, namely the linear programming and the maximum flow computation, and solve the general problem in polynomial time.
Unfortunately, as we will show in this section, the general version of the problem is indeed \NP-hard and thus the approach fails.

Nevertheless, we will show in the next section that, the combination of the techniques provides a tight rounding scheme that gives a $2$-approximation algorithm.

To prove the \NP-hardness (Theorem~\ref{th:hardness}), we use a reduction from a fundamental boolean function satisfactory problem called MAX-DNF.

\begin{figure}[h]
	\centering
	\includegraphics*[width=0.5\textwidth]{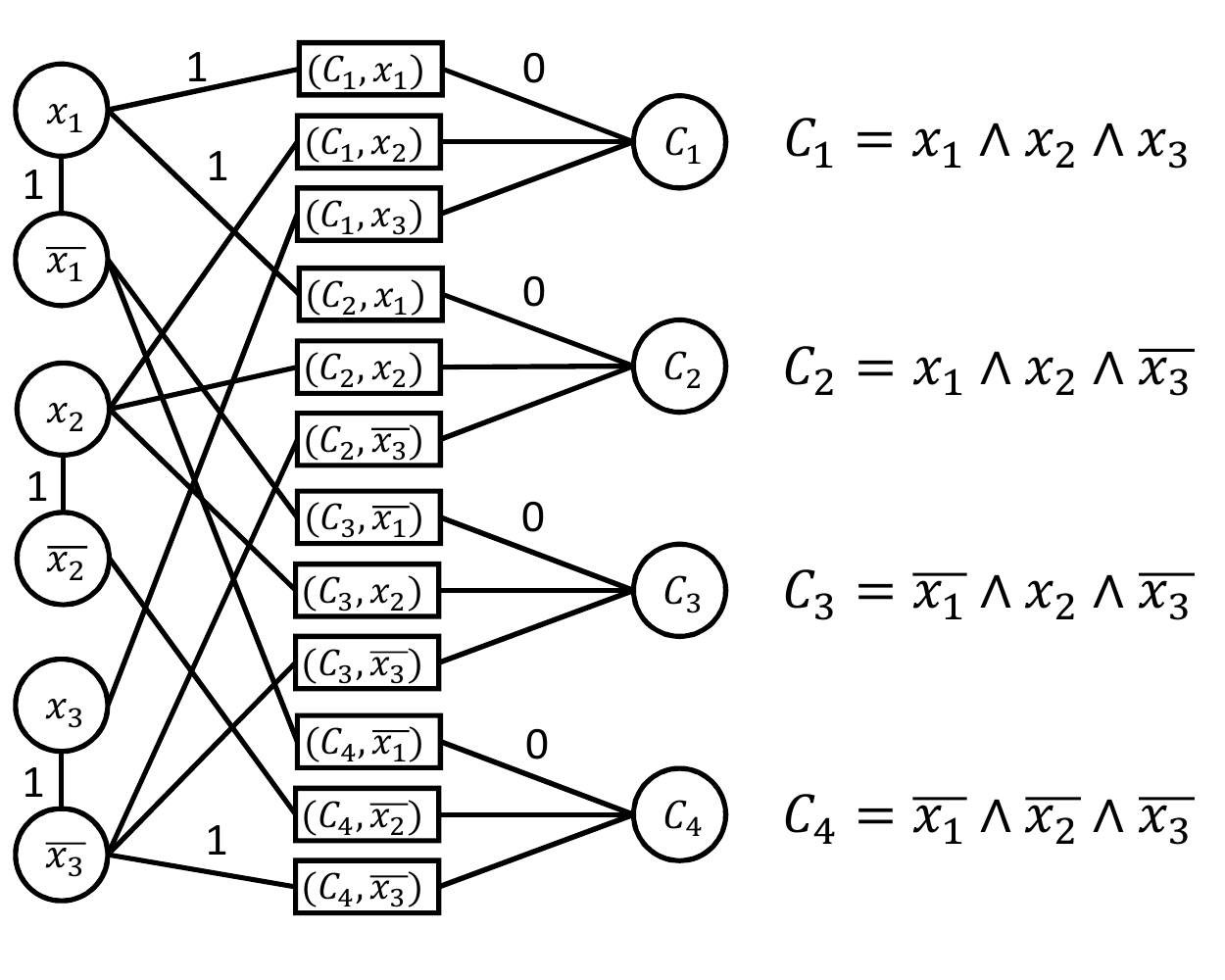}
	\caption{Illustrating example with $p=3$ variables $x_1,x_2,x_3$ and $q=4$ clauses $C_1,C_2,C_3,C_4$.}
	\label{fig:hardness}
\end{figure}

\begin{definition}[MAX-DNF]
	In the  problem, we have boolean variables $x_1,\ldots,x_p$ and clauses $C_1,\ldots,C_q$, where each clause $C_i$ is the conjunction (``and'') of variables or their negations (see Figure~\ref{fig:hardness} for an example).
	The problem aims at finding an assignment to the variables such that a maximum number of clauses are satisfied.
\end{definition}

The problem is shown to be \NP-hard by~\cite{eor/BazganP03,eor/EscoffierP07}.
In the following, we show how to reduce the MAX-DNF problem to ours.
In other words, we show that if the network defending problem can be solved in polynomial time, then the MAX-DNF problem can also be solved in polynomial time.

\paragraph{Reduction.}
Given any MAX-DNF problem instance, we create $2p$ nodes, which are labeled by $x_1,  \overline{x_1},\ldots,x_p,\overline{x_p}$; and $q$ nodes labeled by $C_1,\ldots,C_q$.
We call these nodes \emph{variable nodes} and \emph{clause nodes}, respectively.
Let there be an edge between every pair of $x_i$ and $\overline{x_i}$.
For every clause $C_i = a_1 \wedge \ldots \wedge a_k$, where each $a_j$ represents a variable or its negation, we create $k$ nodes, each labeled by $(C_i,a_j)$.
We call these nodes \emph{connectors}.
Let connector $(C_i,a_j)$ be connected to $a_j$ and $C_i$.
Note that in the resulting graph, every connector has degree two, every variable node has degree equal to its total number of appearances in the clauses, and every clause node has degree equal to the number of variables it contains (see Figure~\ref{fig:hardness} for an illustrating example).

Let $w_e = 1$ for the edges $e$ adjacent to variable nodes; let $w_e = 0$ for the edges $e$ adjacent to clause nodes.
Let $g_u = g'_u = 1$ for variable nodes and clause nodes; let $g_u = g'_u = 0$ for connectors.
For every variable node or connector, let $\ub_u = \lb_u = 1$; for every clause node, let $\ub_u = \frac{1}{q}$ and $\lb_u = 0$.
Let $R = p + \frac{q-t}{q}$, for some $t < q$.

\medskip

We show that there exists a defending strategy with defending result $0$ if and only if there exits an assignment to the variables such that at least $t$ clauses are satisfied.

First, if there exits an assignment to the variables such that at least $t$ clauses are satisfied, then we
\begin{itemize}
	\item allocate $1$ unit of defending resource to every variable node that is ``true'' in the assignment;
	\item allocate $\frac{1}{q}$ defending resource to every unsatisfied clause.
\end{itemize}
Trivially, the total defending resource required is at most $p + \frac{q-t}{q}$. Next we show that the defending result is $0$.

Since the edges adjacent to variable nodes have weight $1$, every variable node $u$ has defending power $1 = \ub_u$.
Moreover, if a variable (or its negation) is true, then each of its connector neighbors $v$ has defending power $1= \ub_v$.
Consequently, if a clause is satisfied, then all its connector neighbors have defending power above their lower bound.
Therefore, the defending result is $0$, as all variable nodes and clause nodes are well-defended.

Next we show the other direction, i.e., the optimal defending strategy corresponds to an assignment of variables such that at least $t$ clauses are satisfied.

We first show that every defending strategy can be transformed into a canonical form, while the defending result is not affected.
Fix any defending strategy $\{ r_u \}_{u\in V}$.

\paragraph{Canonical Transformation.}
For every connector $u$, if $r_u > 0$, then we reallocate the resource to its variable node neighbor.
Since the edge between $u$ and its variable node neighbor has weight $1$ while the edge between $u$ and its clause node neighbor has weight $0$, reallocating the resource does not decrease the defending power of any node.
Next, if $r_u < 1$ for a variable node $u$, then we reallocate its defending resource to its variable node neighbor, which does not change its defending power.
The defending power of the connector neighbors of $u$ will be decreased (to $0$).
However, since $r_u < 1$, the defending powers of these connectors were less than their lower bounds.
Hence decreasing their defending power does not affect the defending result.

\medskip
Fix the optimal defending strategy of canonical form.
Suppose the defending result is $0$.
Then at least one of $x_i$ and $\bar{x}_i$ must be assigned defending resource $1$.
Given that $R<n+1$, exactly one of $x_i,\bar{x_i}$ has defending resource $1$, while the other has defending resource $0$ (which corresponds to an assignment to the variables).
A connector has defending power $1$ if it is connected to a variable node with defending resource $1$.
Hence, if all neighbors (which are connectors) of some clause node $C_i$ have defending power $1$, then we do not need to allocate any defending resource to $C_i$.
On the other hand, if a clause $u$ is not satisfied, then $\frac{1}{q}$ defending resource must be allocated to $u$.
Since the total defending resource deployed is at most $p+\frac{q-t}{q}$, we know that at most $q-t$ clause nodes receive non-zero defending resource.
Therefore, we can retrieve an assignment of variables such that at least $t$ clauses are satisfied given the optimal defending strategy.

Note that by varying $t$ from $1$ to $q$, we can solve the MAX-DNF problem using $q$ computations\footnote{Indeed, we can apply a binary search, which reduces the number of computations to $O(\log q)$.} of our problem.
Thus the problem is \NP-hard.

Since it is \NP-hard to distinguish whether $\opt = 0$ for the above hard instance,
the problem does not admit any approximation algorithm with bounded ratio: any such algorithm can be used to distinguish whether $\opt = 0$.
\begin{corollary}
	The network defending problem (that aims at minimizing the defending result) does not admit any polynomial-time approximation algorithm, unless \P = \NP.
\end{corollary}

\section{Resource Augmentation Algorithms}\label{sec:approx}

Since the network defending problem is not approximable, instead of comparing the gain of the attacker with bounded defending power, we measure the approximation ratio of the problem in terms of defending power deployed in this section.
Formally speaking, we call an algorithm $k$-approximate if by using $R$ defending resource, the defending result is at most that of any optimal defending strategy using $\frac{R}{k}$ resource.
In other words, a $k$-approximate algorithm guarantees that by using $k$ times more resource, the defending result is at least as good as the optimal solution (without augmenting the resource).

In this section, we present a $2$-approximate algorithm for the general network defending problem (Theorem~\ref{th:approx}).

As before, for every fixed $\alpha\in \mathcal{G} = \{g_u:u\in V\}\cup \{g'_u:u\in V\} \cup \{0\}$, we check if it is possible to allocate the $R$ defending resource such that the resulting defending result is at most $\alpha$.
Note that to achieve an approximation ratio of $2$, we show that, as long as $\alpha$ is achievable (by the optimal solution) using $\frac{R}{2}$ defending resource, our algorithm (with $R$ defending resource) computes in polynomial time a defending strategy with defending result at most $\alpha$.

\paragraph{Vulnerable and Crucial Nodes.}
Again, let $A = \{u: g_u > \alpha\}$ be the vulnerable nodes, and $B = \{ u : g'_u>\alpha\}$ be the crucial nodes.
Then the problem is (similar to what we have done in Section~\ref{ssec:poly_isolated}) to
(1) decide a set of nodes $S\subseteq B$;
(2) allocate the resources such that every $u\in S$ has a defending power at least $\ub_u$, and every $v\in A\cup \bigcup_{u\in B\setminus S} N(u)$ has a defending power at least $\lb_v$.

Note that for every fixed $S\subseteq B$, the second step can be easily done using an LP, as we have done in Section~\ref{ssec:poly_lb=ub}.
The difficulty, thus, lies in identifying the subset $S\subseteq B$ such that the required total defending resource is minimized.

\paragraph{Integer Program Formulation.}
Observe that we can formulate the problem into a feasibility integer program.
In the integer program, there is a variable $y_u\in \{0,1\}$ associated with every $u\in B$ indicating whether $u$ is in $S$, i.e., receives defending power $\ub_u$; there is a variable $y_v\in \{0,1\}$ associated with every $v\in V\setminus A$ indicating whether $v$ receives defending power $\lb_v$.
The solution $y$ is feasible if 
\begin{enumerate}
	\item[(1)] for every $u\in B$ such that $y_u = 0$, all its neighbors $v$ in $V\setminus A$ have $y_v = 1$. In other words, $y_u+y_v \geq 1$;
	\item[(2)] there exists $\{ r_u \}_{u\in V}$ with $\sum_{u\in V} r_u = R$ such that
	\begin{align*}
	\lb_u + y_u\cdot (\ub_u-\lb_u) & \leq p_u,\qquad \forall u\in B \\
	\lb_u & \textstyle \leq p_u,\qquad \forall u\in A\setminus B \\
	y_v\cdot \lb_v & \textstyle \leq p_v,\qquad \forall v\in V\setminus A
	\end{align*}
	where $p_u = r_u + \sum_{v\in N(u)} w_{uv}\cdot r_v$ is the defending power of $u$, under defending strategy $\{ r_u \}_{u\in V}$.
\end{enumerate} 

In other words, constraint (1) requires that, for every $u\in B$, either $u$ has defending power $\ub_u$, i.e., $y_u = 1$; or all its neighbors\footnote{Given that nodes $v\in A$ have defending power at least $\lb_v$ in any case, we only need to put constraints on its neighbors in $V\setminus A$.} $v$ has defending power at least $\lb_v$.
Constraint (2) requires that there exists a defending strategy using total resource $R$ such that all nodes receive the specified defending power.
Given that we are aiming for a $2$-approximation, we change the constraint $\sum_{u\in V} r_u = R$ to $\sum_{u\in V} r_u = \frac{R}{2}$, i.e., we are comparing with the optimal defending strategy using $\frac{R}{2}$ defending resource.

The standard LP relaxation (similar to the minimum cut problem) can be formulated as follows.
Let $F := E\cap (B\times (V\setminus A))$ be the edges between $B$ and $V\setminus A$.
\begin{align*}
\min. \qquad\qquad 0 & \\
\text{s.t.}\qquad y_u + y_v &\geq 1, \qquad \forall (u,v)\in F \\
y_u &\geq 0, \qquad \forall u\in B\cup (V\setminus A) \\
\lb_u + y_u\cdot (\ub_u-\lb_u) & \leq p_u,\qquad \forall u\in B \\
\lb_u & \textstyle \leq p_u,\qquad \forall u\in A\setminus B \\
y_v\cdot \lb_v & \textstyle \leq p_v,\qquad \forall v\in V\setminus A \\
\textstyle r_u + \sum_{v\in N(u)} w_{uv}\cdot r_v & = p_u,\qquad \forall u\in V \\
\textstyle \sum_{u\in V} r_u & = R/2.
\end{align*}

It is easy show that the above LP is indeed a relaxation of the integer problem of interest.
As long as constraint~(1) is satisfied, the first constraint of the above LP is satisfied.
The remaining constraints are satisfied by constraint~(2).

Next, we show that the LP is feasible if there exists a defending strategy $\{r_u\}_{u\in V}$ using $\frac{R}{2}$ resource with defending result at most $\alpha$.
Fix any such defending strategy.
It gives a feasible solution for the above LP as follows:
\begin{enumerate}
	\item $\forall u\in B$, set $y_u = 1$ if $p_u \geq \ub_u$; $y_u = 0$ otherwise;
	\item $\forall v\in V\setminus A$, set $y_v = 1$ if $p_v\geq \lb_v$; $y_v = 0$ otherwise.
\end{enumerate}
It suffices to check the first set of constraints to guarantee feasibility.
By the feasibility of the strategy, for every pair of neighbors $u\in B$ and $v\in V\setminus A$, at least one of $y_u, y_v$ is set to be $1$. Thus the constraints are satisfied.

It remains to show that, if the LP is feasible, then our algorithm computes (in polynomial time) a defending strategy with defending result at most $\alpha$ using $R$ defending resource.
Our defending strategy is constructed using any feasible solution of the LP as a guidance.
In the following, we give a geometric interpretation for the solution, which reveals some connections between our approximation algorithm and our max-flow based algorithm in Section~\ref{ssec:poly_isolated}.

\paragraph{Geometric Interpretation.}
Imagine there are two extra nodes $s$ and $t$, where $s$ is at position $0$ and $t$ is at position $1$.
It would be easier to imagine the two nodes as the source and sink, as we have done in Section~\ref{ssec:poly_isolated}.
For every $u\in B$, variable $y_u\in [0,1]$ indicates the distance between $s$ and $u$; for every $v\in V\setminus A$, $y_v$ indicates the distance between $v$ and $t$.
If $y_u = 1$ for some $u\in B$, i.e., we put node $u$ at position of $t$, then we ensure that $u$ has defending power $\ub_u$ (recall that in Section~\ref{ssec:poly_isolated}, this is the case when edge $(u_\text{in},u_\text{out})$ is cut).
If $y_v = 1$ for some $v\in V\setminus A$, i.e., we put node $v$ at position of $s$, then we ensure that $v$ has defending power $\lb_v$ (recall that in Section~\ref{ssec:poly_isolated}, this is the case when edge $(v_\text{in},v_\text{out})$ is cut).
The constraint $y_u + y_v \geq 1$ for every pair of neighbors $u\in B$ and $v\in V\setminus A$ guarantees that the position of $v$ is before that of $u$.
Specifically, if $u\in B$ does not have defending power at least $\ub_u$, then all its neighbors $v\in V\setminus A$ should have defending power at least $\lb_v$.

\paragraph{Rounding and Feasibility.}
Given the optimal solution $(y,r)$ for the above LP, we construct a feasible integral solution $(Y,r')$, i.e., a defending strategy using $R$ defending resource, as follows.
For every $u\in B\cap (V\setminus A)$, set $Y_u = 1$ if $y_u\geq \frac{1}{2}$; $Y_u = 0$ otherwise.
Set $r'_u = 2\cdot r_u$.
Observe that after the rounding, all variables $Y$ take values in $\{0,1\}$, and the total defending resource used is $R$.

For the first set of constraints, observe that for every pair of neighbors $u\in B$ and $v\in V\setminus A$, (by feasibility of $y$) at least one of $y_u, y_v$ is at least $\frac{1}{2}$.
Thus after rounding at least one of them is $1$.
In other words, the integral solution satisfies the first set of constraints.
For the third, fourth and fifth sets of constraints, observe that our integral solution increases the defending power $p_u$ of every node $u$ by a factor of $2$, while increases $y_u$ by a factor of at most $2$. Thus these constraints are all satisfied.

As long as it is possible to achieve defending result $\alpha$ using $\frac{R}{2}$ defending resource, our algorithm (which uses $R$ defending resource) computes a feasible defending strategy in polynomial time. Thus our algorithm is $2$-approximate, which completes the proof of Theorem~\ref{th:approx}.

\paragraph{Integrality Gap.}
While we do not have a matching lower bound on the approximation of the problem, we show that any rounding algorithm based on this LP cannot do better than $2$-approximate.
More specifically, there exists an instance for which any solution achieving defending result $\alpha$ requires defending resource $R$, while there exists a fractional solution for the above LP (using $\frac{R}{2}$ resource) that is feasible.
In other words, the integrality gap of the LP relaxation is $2$.

Let the graph be two nodes $u$ and $v$ connected by an edge.
Let $R = w_{uv} = g_u = g'_u = 1$ and $g'_v = g_v = 0$.
Let $\lb_u = 0$ and $\ub_u = 1$; $\lb_v = 1$ and $\ub_v = 2$.
In the optimal solution, by allocating $1$ defending resource arbitrarily, the attacker has gain $0$.
Moreover, if $R < 1$, then no matter how the defending resource is distributed, the attacker always gains $1$ by attacking $u$.
However, in the fractional solution, by setting $y_u = y_v = r_u = 0.5$ and $r_v = 0$, the solution is actually feasible for the above LP.
In other words, the LP is feasible when $R = 0.5$.

\section{Conclusion}
In this paper, we propose a network security game that allows the sharing of defending resource between neighbor nodes, and the spread of attack damage to the neighbors of the target.
The model captures features of many real-world applications that are not covered by existing network security game models.
We present polynomial time algorithms for two natural and important special cases of the problem.
We show that the general problem is \NP-hard, and propose an LP-rounding based $2$-approximation algorithm.

The most interesting open problem is whether the approximation ratio $2$ we obtained in Section~\ref{sec:approx} can be improved.
While we have shown that our rounding is tight, the integrality gap does not directly translate to hardness result on inapproximability.
We believe that it is possible to prove an \APX-hard result for the general problem.

\section*{Acknowledgement}

Minming Li was partially supported by NNSF of China under Grant No. 11771365, and by Project No. CityU 11200518 from Research Grants Council of HKSAR.

This research was sponsored by the U.S. Army Research Laboratory and the U.K. Ministry of Defence under Agreement Number W911NF-16-3-0001. The views and conclusions contained in this document are those of the authors and should not be interpreted as representing the official policies, either expressed or implied, of the U.S. Army Research Laboratory, the U.S. Government, the U.K. Ministry of Defence or the U.K. Government. The U.S. and U.K. Governments are authorized to reproduce and distribute reprints for Government purposes notwithstanding any copyright notation hereon. In addition, Tran-Thanh and Li would like to acknowledge the financial support from the Royal Society’s Kan Tong Po Fellowship (KTP \textbackslash R1 \textbackslash 170018). 

The research leading to these results has received funding from the European Research Council under the European Community’s Seventh Framework Programme (FP7/2007-2013) / ERC grant agreement No. 340506.

{
	\bibliography{ADGame}
	\bibliographystyle{alpha}
}

\end{document}